\documentclass[manuscript]{acmart}

\setcopyright{acmlicensed}
\copyrightyear{2025}
\acmYear{2025}
\acmDOI{XXXXXXX.XXXXXXX}
\acmConference[CHI '25]{the CHI 2025 Workshop on Technology Mediated Caregiving for Older Adults Aging in Place}{April 26 - May 1, 2025}{Yokohama, Japan}

\begin{document}

\title{Navigating Privacy and Trust: AI Assistants as Social Support for Older Adults}

\author{Karina LaRubbio}
\email{karina_larubbio@brown.edu}
\orcid{0000-0003-2514-7546}
\affiliation{
  \institution{Brown University}
  \city{Providence}
  \state{Rhode Island}
  \country{USA}
}

\author{Malcolm Grba}
\email{malcolmgrba@g.harvard.edu}
\orcid{0009-0000-6775-0314}
\affiliation{%
  \institution{Harvard University}
  \city{Cambridge}
  \state{Massachusetts}
  \country{USA}
}

\author{Diana Freed}
\email{diana_freed@brown.edu}
\orcid{0000-0001-5262-4147}
\affiliation{
  \institution{Brown University}
  \city{Providence}
  \state{Rhode Island}
  \country{USA}
}

\renewcommand{\shortauthors}{LaRubbio et al.}

\begin{abstract}
 
AI assistants are increasingly integrated into older adults’ daily lives, offering new opportunities for social support and accessibility while raising important questions about privacy, autonomy, and trust. As these systems become embedded in caregiving and social networks, older adults must navigate trade-offs between usability, data privacy, and personal agency across different interaction contexts. Although prior work has explored AI assistants' potential benefits, further research is needed to understand how perceived usefulness and risk shape adoption and engagement. This paper examines these dynamics and advocates for participatory design approaches that position older adults as active decision makers in shaping AI assistant functionality. By advancing a framework for privacy-aware, user-centered AI design, this work contributes to ongoing discussions on developing ethical and transparent AI systems that enhance well-being without compromising user control.
\end{abstract}

\keywords{Aging in place; older adults; AI assistants; privacy; social support}

\maketitle

\section{Introduction}

As the population of adults over 65 years of age in the United States is projected to nearly double by 2060 \cite{vespa_demographic_2018}, the need for innovative solutions to support aging in place has become increasingly urgent. While many older adults prefer to age in place, concerns about social isolation and loneliness persist, even in community environments. Technology has long played a role in facilitating remote communication, and the recent emergence of generative artificial intelligence (AI) introduces new possibilities for scalable AI assistants designed to provide companionship and social support \cite{broadbent_elliq_2024, wong_voice_2024}.

Prior research underscores the link between social participation and improved health outcomes \cite{dawson-townsend_social_2019, kullgren_national_2023, noauthor_social_2021}, reinforcing the potential of AI-driven tools to foster meaningful connections. However, the introduction of AI assistants to older adults aging in place raises complex questions about privacy, security, and ethical implications. Unlike traditional communication technologies, AI assistants actively engage in social interactions, learn user preferences, and adapt their responses over time. This dynamic interaction model presents unique challenges in ensuring transparency, safeguarding sensitive information, and maintaining user autonomy.

Although best practices for AI assistants that support older adults' physical health have been explored \cite{xie_fitpal_2024, yang_talk2care_2024, brewer_empirical_2022, brewer_if_2022, so_they_2024}, more work is needed to understand the functionality and privacy preferences of AI assistants designed for social support in various usage scenarios. In private spaces, AI assistants may engage in one-on-one interactions. However, these interactions may become visible in ways that users may not expect in communal settings such as gatherings and independent living facilities. This presents unique challenges for AI assistants to dynamically support collaborative functionality while maintaining individual users' privacy. 

We investigate AI assistants that promote social interaction and well-being without compromising safety. Our research explores how AI assistants are integrated into the daily lives of older adults and become entangled in both public and private spaces. We critically evaluate their privacy and safety implications, along with opportunities to communicate these risks to older adults, to encourage safer, more informed technology adoption.

\section{Research Background}

Older adults may experience feelings of loneliness and social isolation, which are associated with poor health outcomes. The 2023 National Poll on Healthy Aging in the United States found that one in three older adults reported feelings of isolation among a nationally representative survey of 2,563 respondents \cite{kullgren_national_2023}. Additionally, the survey found associations between high measures of loneliness and poor mental and physical health. The World Health Organization identified social isolation and loneliness as important social determinants of health globally, which impact both physical and mental well-being for older adults \cite{noauthor_social_2021}. Even among community-dwelling older adults, 24\% of 6,649 respondents to the 2011 National Health and Aging Trends Study in the United States characterized themselves as socially isolated \cite{cudjoe_epidemiology_2020}. 

Prior work has identified the role of informal caregivers, such as adult children and other family members, in providing social support to older adults. Due to demographic trends, many informal caregivers balance the responsibilities of caring for children alongside aging parents as part of the "sandwich generation" \cite{juliana_menasce_horowitz_more_2022}. Informal caregivers seek support for many practical and emotional challenges associated with their responsibilities. In an analysis of a Reddit community for caregivers of older adults, it was found that informal caregivers sought support for their feelings of guilt due to aging care recipients' loneliness \cite{homaeian_community_2025}. Previous work has called for interventions to provide support for informal caregivers due to the often negative impacts of caregiving on their health and well-being \cite{schorch_designing_2016, bom_impact_2018}. 

Thus, along with interventions to support informal caregivers, there is an opportunity to use technology to provide additional social support to older adults. During the COVID-19 pandemic, digital technologies such as video calling presented opportunities for older adults to connect with their social support networks despite physical accessibility barriers, although in-person interactions remained highly valued \cite{kelly_more_2024, wang_understanding_2024}. 
Although digital technologies have been proposed and evaluated to help older adults access new and existing social support networks, there is an opportunity to develop more support solutions that align with older adults' preferences and scalability needs.

\subsection{AI Assistants}

A technology of particular interest to our research is AI assistants. We expand on the definition of AI assistants as AI-driven platforms and devices that can be used for social support. These include general-purpose voice assistants, such as Apple's Siri and Amazon's Alexa. AI-driven devices and platforms with other interaction modalities, such as text and other non-verbal communications, are also considered. Additionally, we include AI-driven tools designed specifically for older adults with various interface styles, such as animatronic pets or robotic assistants. 

Increasingly, AI assistants are being designed, evaluated, and deployed to provide social support as companions to older adults. Some evaluations focus on the ways that general-purpose AI assistants, such as Amazon Alexa, Google Home, and ChatGPT, may be used by older adults for social support in various domains, ranging from health advice \cite{brewer_if_2022, harrington_its_2022, brewer_empirical_2022} to reminiscence \cite{jin_exploring_2024, zhai_exploring_2024, cuadra_designing_2023}. Other explorations take a speculative approach to designing specified systems to support older adults' social support needs, such as participatory and co-design workshops with older adults which aim to define their unique system requirements \cite{so_they_2024, wong_voice_2024, choi_together_2023}. 

Some AI assistants specifically designed to combat loneliness among older adults are commercially available with variable interface designs, including ElliQ, a voice-controlled robot with interactive screen \cite{broadbent_elliq_2024}. ElliQ actively engages users in social interactions and uses an AI algorithm to personalize content to individual users. During beta testing, users reported feeling less lonely when they were with the robot. Other proposed interface designs for older adults assistants include humanoid robots and animatronic pets \cite{portacolone_ethical_2020}. 

\subsection{Design and Functionality Considerations from Older Adults to Promote Agency}
Given their potential utility in providing social support, previous work has evaluated older adults' design and functionality preferences for assistants, some of which integrate AI. Among the general population, the anthropomorphic qualities of AI assistants are often associated with enhanced perceived empathy and positive user experiences \cite{hu_touch_2018, svenningsson_artificial_2019, jensen_trust_2021}. However, older adults often demonstrate different preferences. Coghlan et al. interviewed older adults for their perspectives on various interface designs for robot assistants, including a voice companion, a toy-like vehicle, and an animatronic dog \cite{coghlan_dignity_2021}. Toy-like assistants were sometimes viewed as infantilizing, although participants' varying personalities led to diverse interface preferences. Although the voice companion and vehicle were considered to offer unique functional opportunities, older adults generally praised the control and empathy offered by the robotic pet. Thus, older adults' desire for feelings of control when interacting with assistants exemplifies the importance of considering the unique needs of this population in design processes. 

Interaction functionality constraints may also impact older adults' sense of autonomy and control with AI assistants, particularly in dialogues. Zubatiy et al. suggested that the trial-and-error learning style of general-purpose AI assistants conflicts with the preference of older adults for errorless learning by analyzing qualitative interviews with older adults and their caregivers along with logs of their interactions with a conversational agent \cite{zubatiy_i_2023}. When an error is made on the prompt, the user must restart the interaction without clear feedback on what went wrong. This led some older adult participants to engage in error-avoidant behaviors that complicated their interactions and their ability to learn to use the system, thus diminishing their control. 
Older adults also express preferences for AI assistants to learn about their personalities and preferences in contexts such as health \cite{brewer_empirical_2022, zubatiy_i_2023} and reminiscing \cite{zhai_exploring_2024}. However, human-like conversational styles were perceived as inauthentic and uncanny in some contexts, such as mental health \cite{wong_voice_2024}. In addition to physical design preferences, older adults have diverse preferences for qualities of dialogues, which may vary between application contexts and individual users.

\subsection{Privacy and Well-being Considerations from Older Adults}
Older adults have expressed various concerns about maintaining agency over their privacy and well-being while using AI assistants, especially in the context of social support. Previous work has identified older adults' desire for AI assistants to learn about them, including their personal preferences \cite{zhai_exploring_2024} and health experiences \cite{brewer_empirical_2022, zubatiy_i_2023}. Further, older adults recognize opportunities to use AI assistants to access support for stigmatized topics like drug use, sex, isolation, and dementia that they may be uncomfortable to discuss with healthcare providers, family members, or peers \cite{so_they_2024}. Without appropriate data protections, the collection of especially sensitive user data may pose a privacy risk. Older adults also identify a desire for technologies to consider non-verbal conversational cues to enhance usability \cite{baker_interrogating_2019, cuadra_inclusion_2022}. However, integrating non-verbal cues may require processing and storing data from visual or haptic sensors. Opportunities to enhance AI assistants for personalized support and natural-feeling user interaction appear to require the collection of more user data, creating potential tension between usability and privacy.

In the context of social support, AI assistant interactions may take place in private or public settings, producing implications for privacy and functionality. In an observational study of the usage of a general-purpose voice assistant by older adults in a public setting, it was noted that these systems are not equipped to handle interactions with multiple users simultaneously \cite{cuadra_inclusion_2022}. This led collaborating participants to engage in side conversations hidden from the AI assistant. Alongside functionality challenges, privacy concerns are introduced by the public usage of AI assistants within older adults' immediate social support networks. In a co-design study on voice assistants for health, older adults identified the need to control whether AI assistants share sensitive personal information they may have learned during prior private interactions, depending on whether an untrustworthy person is nearby \cite{so_they_2024}. Considering the community contexts that older adults occupy while aging in place is essential for understanding how AI assistants interact with other components of their social support networks. 

Within older adults' social support networks, there are also concerns about how AI assistants for social support may share information gained during interactions with relevant stakeholders, such as caregivers and family members. Chang et al. showed older adults storyboards of various information-sharing scenarios in the context of AI assistants for caregiving \cite{chang_dynamic_2024}. Participants expressed that the AI assistant's affiliation should be dynamic throughout aging, preserving the older adults' agency for as long as possible and sharing more information in the event of a health decline. 

Beyond immediate support networks, older adults express concern about the lack of transparency surrounding data privacy with AI assistants and AI technologies in general \cite{shandilya_understanding_2022}. Additionally, data privacy policies can be difficult for older adults to find and understand \cite{karen_bonilla_older_2020}. Older adults have expressed concerns about the surveillance of sensitive information collected by AI assistants and suggested sharing data on a need-to-know basis \cite{so_they_2024}. In many aspects, increasing the usefulness and usability of AI assistants requires older adults to disclose more personal information, allowing privacy risks within and beyond their social support networks. 

Older adults express usability and reliability needs to ensure that their well-being is supported with AI assistants intended for social support. For example, communication breakdowns due to the incompatibility of conversational styles between AI assistants and older adults' expectations may hinder their usage of a system \cite{cuadra_designing_2023, zubatiy_i_2023, harrington_its_2022}. Beyond usability, older adults may experience safety concerns while interacting with AI assistants, particularly in sensitive contexts such as emotional and health support. Through interviews, older adults suggested that AI assistants should only be used for low-risk questions in health contexts \cite{harrington_trust_2023}. This conveys that older adults may have a sense of distrust toward AI assistants, a sentiment that may be prohibitive to the use of social support AI assistants without appropriate safeguards. Recent reports of users, mostly adolescents, causing harm to themselves after building personal relationships with AI assistants \cite{roose_can_2024} raise concerns about how these technologies may impact older adults aging in place.

\section{Future Work}
Further research is needed to explore how older adults balance the benefits of AI assistants for care and social support against concerns related to privacy, security, and well-being across different use contexts. Frameworks have been proposed for understanding factors influencing older adults' cybersecurity awareness \cite{morrison_recognising_2023} and technology usage while aging in place \cite{peek_older_2016} separately. However, within the context of emerging technologies such as AI assistants, the interplay of perceived usefulness and perceived risk in older adults' decision-making processes is not well understood.

The increasing demand for personalized interactions with AI assistants raises critical privacy considerations. Given the diverse contexts in which these systems are deployed, it is imperative to investigate best practices that safeguard user privacy while maintaining system usability in both individual and group interactions. AI assistants frequently function within broader social support networks, necessitating careful attention to the contextual appropriateness of information-sharing, particularly with caregivers \cite{chang_dynamic_2024}. Future research should continue to explore how these systems can be seamlessly integrated in ways that align with user expectations and preserve agency, ensuring that personalization does not come at the expense of privacy.

Given older adults' concerns about data management and transparency \cite{karen_bonilla_older_2020, shandilya_understanding_2022}, existing and emerging AI technologies should clarify their data privacy policies in accessible language and, ideally, design secure and privacy-aware systems. Furthermore, possible threats to well-being, which have previously been observed in other vulnerable populations when using AI assistants \cite{roose_can_2024}, demonstrate the importance of guardrails to preserve user safety in sensitive contexts like social support. The protections for data privacy and well-being that AI assistants implement must be clearly communicated to allow older adults and their caregivers to make informed decisions about integrating these technologies into their social support networks. Steps should be taken to understand the best practices for communicating the risks and precautions associated with various AI assistants. 

Therefore, there is an opportunity for the HCI and CSCW communities to examine the role of AI assistants in social support for older adults in terms of privacy, usability and their interaction. Understanding the contexts in which older adults are aging in place will enable a more holistic understanding of preferences for AI assistant functionality and design. Following the fruitful model of previous work in this space \cite{so_they_2024, wong_voice_2024, choi_together_2023, martin-hammond_engaging_2018}, participatory and codesign methods should be used to continue to include the perspectives of older adults in the decision-making process about whether and how to integrate emerging technologies into their social support networks.

\bibliographystyle{ACM-Reference-Format}
\bibliography{bib}

\end{document}